\documentclass[twocolumn,preprintnumbers,amsmath,amssymb]{revtex4}
\usepackage{mathrsfs}
\usepackage{amssymb}
\usepackage{graphicx}
\usepackage{dcolumn}
\usepackage{bm}
\usepackage{textcomp}
\usepackage{amsmath}
\usepackage[titletoc]{appendix}

\newcommand\ket[1]{\ensuremath{|#1\rangle}}
\newcommand\bra[1]{\ensuremath{\langle#1|}}
\newcommand\iprod[2]{\ensuremath{\langle#1|#2\rangle}}

\newcommand\tr{\mathop{\rm tr}\nolimits}
\newcounter{RomanNumber}

\begin{document}
\title{Encoding-side-channel-free and measurement-device-independent quantum key distribution}
\author{Xiang-Bin Wang$ ^{1,2,4} $
\footnote{email: {xbwang@mail.tsinghua.edu.cn}}, Xiao-Long Hu$ ^{1}$, and
 Zong-Wen Yu $ ^{3}$ }
\affiliation{ \centerline{$^{1}$State Key Laboratory of Low
Dimensional Quantum Physics, Department of Physics,} \centerline{Tsinghua University, Beijing 100084,
 China}
\centerline{$^{2}$ Synergetic Innovation Center of Quantum Information and Quantum Physics, University of Science and Technology of China}
\centerline{  Hefei, Anhui 230026, China
 }
\centerline{$^{3}$Data Communication Science and Technology Research Institute, Beijing 100191, China}
\centerline{$^{4}$ Jinan Institute of Quantum technology, SAICT, Jinan 250101,
People¡¯s Republic of China}}
\begin{abstract}
We present  a simple protocol where Alice and Bob only needs sending out a coherent state or not-sending out a coherent state to Charlie. There is no bases switching. We show that this protocol is both encoding-state-side-channel free to the source part and measurement-device-independent. We don't have to control exactly the whole space state of the light pulse, which is an impossible task in practice. The protocol is immune to all adverse due to encoding-state imperfections  in side-channel space such as the photon frequency spectrum, emission time, propagation direction, spatial angular moment, and so on.  Numerical simulation shows that our scheme can reach a side-channel-free result for quantum key distribution over a distance longer than 200 km given the single-photon-interference misalignment error rate of $20\%$.
\end{abstract}


\maketitle

{\em I Introduction}
Guaranteed by principles of quantum mechanics, quantum key distribution (QKD) can provide a secure key for private communication~\cite{BB84,GRTZ02} even though Eve can completely control the channel. However,  there are side-channel effects due to the device imperfections in practice\cite{ILM,H03,wang05,LMC05,wangyang,njp,PNS,PNS1}. In general, we can divide the whole space of an encoding state from a real source into two subspaces, the operational space and the side channel space. Even though the source state encoding looks perfect when it is examined in the operational space, there could be security loopholes due to imperfections in the side channel space, this is what we called side-channel effects. For example, in the BB84 protocol\cite{BB84}, even  though a perfect single-photon source is applied, there are still some side-channel effects which can undermine the security assumed  in the operational space. There could be basis-dependent synchronization errors in pulse emitting time or frequency spectrum difference for different encoding states or bases and Eve can make use of this to judge the basis or encoding state chosen in a certain time window.  In general, all encoding states from the source
  live in an infinite dimensional space which is formed by the operational space and the side-channel space. Though we only use the encoding space  (e.g., polarization) for QKD, Eve can do his attack in side-channel space such as frequency space to obtain information. Given the inevitable imperfections in the side-channel space for the encoding states,  Eve can make use of these and  obtain information without causing any change to  the
  Alice and Bob's observation outcome in the operation space. As we shall show latter, given a lossy channel, by taking side channel attacks to the source, Eve can actually almost obtain full information of a QKD result without disturbing the encoding states in the operational space.  Here we propose a QKD scheme that is both source encoding-state-side-channel-free and measurement device independent (MDI)\cite{curty1,ind3,wang10}.  Although  some existing protocols can also achieve the goal of side-channel-free security~\cite{ind1,ind3}, our protocol presented here is the only one that bases on the matured existing technologies  and there is no demanding on the local detection efficiency.

  Our protocol is {\em not} source device independent. The so called  encoding-state-side-channel free means that we don't worry about any information leakage from the side-channel space of the encoded states. We do have conditions in the operational space. In this work we assume that the encoding states from the source behave perfectly when they are measured in the operational space, but can be imperfect in the side-channel space. Hopefully, the condition on the operational space can be loosened in the future. Actually, there are lots of studies on security with inexact encoding states\cite{wangyang,njp,wang10,ind2} in the operational space. On the other hand, doing calibration only in the operational space seems to be much easier than that in the whole space, which seems to be an impossible task in practice.
  Also, our protocol is side-channel free for the encoding states only. It assumes no  information leakage due to other hidden accesses beyond the sent-out encoding states. Say, the source has no hidden classical information leakage.
  Note that in the existing device-independent protocols, there are also similar assumed conditions for security, although the conditions are on the measurement devices rather than the encoding states. Since our protocol is measurement device independent, there is no condition on the measurement devices for security. The security of our protocol is obviously stronger than the normal MDI-QKD because there is no security loophole in the encoding states of our protocol. Compared with the existing device-independent QKD, our protocol has a drawback in that it is not entirely source-independent, but it has an advantage of measurement-device-independent. Note that the device independent QKD is not MDI, because it requests no information leakage of the measurement outcome. Most importantly, our protocol only rely on matured technologies and it can achieve a meaningful secure distance, a distance longer than 200 kilometers even though the misalignment error rate is as large as 20\%.

{\em II Some side-channel attacks and  Theorem 1.}  Consider a two-basis QKD protocol, such as the BB84 protocol, where there are $X$ basis and $Z$ basis in the protocol.
Suppose we take state-encoding in the photon polarization space and we regard the polarization space as the operational space. In the existing methods in generating the different encoding states, we need either use different diodes to generate different encoding states or use only one diode together with randomly chosen modulations, such as flipping,  rotation, phase shift, etc.  These operations can cause differences  in the side channel space.
In particular, the frequency spectrums can be a little bit different for different encoding states or  different bases. In principle, by detecting frequency difference, Eve has a chance to know which encoding state is used in the operational space  and she will cause no change to the states in operational space. As another example, if different encoding states  are actually emitted at different  time, Eve may just measure the photon with a very precise clock and she can sometimes know the coding state almost exactly if the photon wave packet collapses at certain time intervals. Also, Eve may make use of the channel loss, she can choose to block all those photons on which the  side-channel attack done by her is not successful. Thus, small bias of a state in the side-channel space may flaw the whole protocol.

 Fortunately, the ideal source is not the only secure source. A real-life source is secure if it can be  mapped from  an ideal source.  Suppose in a certain protocol $\mathcal K$ requests $k$ different encoding states. For example, in the BB84 protocol, we need 4 encoding states. Perfect source $\mathcal P$ always produces a perfect encoding state in every time window and all states are identical in the side channel space. An imperfect source produces non-ideal encoding states in the whole space, with imperfections in the side-channel space and the states in side-channel space can change from time to  time. Say, at a certain time window, the imperfect source produces the $k$ encoding states in the whole space and we name them as set $\mathcal S_i$. We then randomly choose one from $\mathcal S_i$ and send it out for QKD.
  If there is always a (time-dependent) quantum process $\mathcal M_i$ that maps the $k$ perfect states to the corresponding $k$ non-ideal states in set $\mathcal S_i$, then  we say source $\mathcal S$ can be  mapped from a perfect source $\mathcal P$.
\\{\bf Theorem 1} In any QKD protocol, a real-life  source $\mathcal {S}$ emitting imperfect states in the whole space is equivalent to a perfect (virtual) source $\mathcal P$ emitting perfect states which are all identical in side channel space  if there exists a quantum process $\mathcal M$ that can map source $\mathcal P$ to source  $\mathcal S$. The final key of a QKD protocol using source $\mathcal {S}$ can be calculated by assuming that the virtual source $\mathcal P$ were used.

\begin{figure}
    \includegraphics[width=250pt]{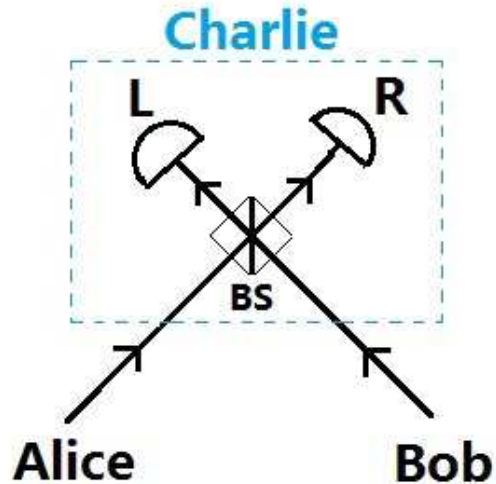}
    \caption{Schematic picture of side-channel-free QKD.}\label{fig:1}
    \end{figure}

This conclusion is rather obvious.  Suppose $\mathcal S$ is insecure, then in a QKD protocol where the ideal source $\mathcal P$ is applied, Eve can first use the quantum process $\mathcal M$ to transform it into  source $\mathcal S$ and then attack the QKD protocol as if the protocol used source $\mathcal S$. This means that if $\mathcal S$ is not secure then $\mathcal P$ is not secure either.
Here in the theorem above, the quantum process $\mathcal M$ is not limited to a unitary process, although in showing the side-channel free property  of our protocol
we mainly use a unitary map.
\\{\em III Our protocol}
\\In our protocol, we use the idea of twin-field QKD (TFQKD)\cite{nature18} with the  classical bit value encoding done by  the decisions of {\em sending}
or {\em not-sending} made by Alice and Bob. The schematic picture of side-channel-free QKD is shown in Fig.~\ref{fig:1}.
 Our protocol is different from \cite{wxb} in that we directly use the whole non-random-phase coherent states.
There are 3 parties, Alice, Bob, and the third party Charlie\cite{nature18}. {\em They} (Alice and Bob) will use coherent states and vacuum only.
 We first present our protocol in the operational space only and then show why it is side-channel free using our {\em Theorem} 1.
\\Protocol R, real protocol
\\ R-1 At any time window $i$, Alice (Bob) always prepare a coherent state $|\alpha_A\rangle=\sum_{n=0}^\infty\frac{e^{-\mu/2} {\mu}^{n/2}e^{in\gamma_A}}{\sqrt{n!}}$ ($|\alpha_B\rangle=\sum_{n=0}^\infty\frac{e^{-\mu/2} {\mu}^{n/2}e^{in\gamma_B}}{\sqrt{n!}}$) and announces the states, including the global phases $\gamma_A,\gamma_B$.
 With a probability $q$ Alice (Bob) decides {\em sending}, and with a probability $1-q$ she (he) decides {\em not-sending}. If she (he) decides {\em sending}, she (he) sends  out the coherent state $|\alpha_A\rangle$  ($|\alpha_B\rangle$)  to Charlie and puts down a classical bit value $1$ ($0$) locally; if she (he) decides {\em not-sending} she (he) does not send out anything, i.e., she (he) sends out a vacuum to Charlie and puts down a classical bit value $0$ ($1$) locally.
\\ R-2 Charlie announces his measurement outcome and hence determines the {\em effective events}: an event with one and only one detector clicking announced by Charlie.
A time window or classical bit value corresponding to an effective event is named as an effective time window or effective bit.
\\ \underline{Definition} $\tilde Z$-window: A time window when Alice decides {\em sending} and Bob decides {\em not-sending}, or Alice decides {\em not-sending} and Bob decides {\em sending}.
\\ R-3
  Through classical communication, {\em they} take a random subset of time windows $v$ to test the bit-flip error rate; {\em they} take another random subset of time windows $u$ to verify the upper bound value of phase-flip error rate $\bar e^{ph}$ of the  effective bits corresponding to $\tilde Z$-windows, those time windows when one party from Alice and Bob decides {\em sending} and the other party decides {\em not-sending}. {\em They} discard effective bits from $v$-windows or $u$-windows after error test.
 \\R-4 {\em They} distill (by taking error correction and privacy amplification to ) the remaining effective bits after error test,  with the asymptotic result for number of final bits:
 \begin{equation}\label{kr00}
 n_F = n_{\tilde Z}-n_{\tilde Z}H(\bar e^{ph})- f n_t H(E_Z)
 \end{equation}
 where the entropy functional $H(x)=-x\log_2 x -(1-x) \log_2 (1-x)$; $n_{\tilde Z}$  ($n_t$) is the number of remaining  effective bits from $\tilde Z$-windows (all time windows) after error test; $f$ is the correction efficiency factor.
\\ \underline{Note 1} The encoding is done by decisions on {\em sending } or {\em not-sending} made by Alice and Bob. More precisely, the {\em sending} or {\em not-sending} decision of a time window  always corresponds to the local classical bits $1$, $0$ to Alice, or 0, 1 to Bob. We can also imagine that whenever Alice (Bob) decides {\em sending} or {\em not-sending}, she (he) always produces a local ancillary photon-number state $|1\rangle$ or $|0\rangle$ and the corresponding bit values are encoded in the local ancillary state. To Alice (Bob), state $|0\rangle$ corresponds to a bit value 0 (1) and state $|1\rangle$ corresponds to a bit value 1 (0). This is equivalent to say that {\em they} (Alice and Bob) have used an extended state including real-photon state which will be sent out to Charlie and ancillary state placed locally. For example, in a certain window when Alice decides sending and Bob decides not sending, we can imagine that {\em they} have actually prepared an extended state
\begin{equation}
\left(\rho_{A}\cdot |0\rangle\langle 0|\right)\otimes |10\rangle\langle 10| .
\end{equation}
where $\rho_A=|\alpha_A\rangle\langle\alpha_A|$. The real-photon state $\rho_A\cdot  |0\rangle\langle 0| $ will be sent out to Charlie. In this paper, we shall name the state left to the tensor-product symbol $\otimes$ as the real-photon state, and the state right to the tensor-product symbol $\otimes$ as the ancillary-photon state. As stated already, each one's bit value is actually encoded in the local ancillary photon-number state.
\\\underline{Note 2} The bit-flip error and phase-flip error. Bob makes a wrong bit  encoding if his bit value is different  from Alice's bit value at a certain time window.  To every event in the subset $v$ randomly taken by {\em them}, {\em they} announce each one's bit values (decision on {\em sending} or {\em not-sending}) to judge the bit-flip error rate. Also, as shown in the Appendix, using the details of events in set $v$, {\em they} can calculate the phase-flip error rate $\bar e^{ph}$
by Eq.(\ref{phasef}).
\\ \underline{Note 3} Charlie's compensation. To obtain a good key rate, {\em they} need a low phase-flip error rate. In the protocol, an event of right-detector clicking due to the real-photon state $|\alpha_A\rangle\otimes |\alpha_B\rangle=|\sqrt\mu e^{i\gamma_A}\rangle\otimes|\sqrt\mu e^{i\gamma_B}\rangle $ will contribute to the phase-flip errors. The states $|\alpha_A\rangle$ and  $|\alpha_B\rangle$ are publicly announced, Charlie can do compensation to remove the global phases $\gamma_A$ and $\gamma_B$ in the states, so that the clicking detector is very unlikely to be the right detector given the real-photon state $|\alpha_A\rangle\otimes |\alpha_B\rangle$. Very importantly, as shown in the Appendix, although Charlie's collaboration can lead to a high key rate, the security of the protocol does not rely on Charlie's honesty.
\\Note 4 Why it is side-channel free ?
Intuitively speaking, our protocol takes no physical-bases switching.
Other protocols, such as the BB84, MDIQKD protocol, the TFQKD, and so on, they all need to take modulation to the outcome states differently  in switching between  different bases. We can give a strict proof for the side-channel free property of our protocol. Here we show directly by our Theorem 1 that the protocol R is side channel free, say, if it is secure in operational space, it must be also secure in the whole space including the side channel space.
Later in the appendix we present the measurement-device-independent security proof of the protocol itself in operational space only, i.e., in the case we don't consider any side-channel effects.
  Here we only need to show that the protocol is the encoding-state-side-channel free.
 In step R-1, {\em they} try to prepare a coherent state $|\alpha_A\rangle$, $|\alpha_B\rangle$ in the operational space.
 In the  ideal case we don't need to consider the side-channel space.  For example, in the case that any Fock state $|n\rangle$, no matter it is from Alice's state $|\alpha_A\rangle$ or from Bob's state $|\alpha_B\rangle$, are identical in side-channel space.  So, we shall just use the original notations of $|\alpha_A\rangle$ and $|\alpha_B\rangle$ to represent the {\em ideal whole-space state}.
 However, what is actually prepared in a real experiment must  have imperfections in the side-channel space.
But {\em the vacuum state has no side channel space} therefore we only need to consider side-channel-space for the non-vacuum parts in each coherent state.
Say, $|\alpha_x\rangle=e^{-\mu/2}|0\rangle + \sqrt{1-e^{-\mu}}|\tilde \alpha_x\rangle$and $\sqrt{1-e^{-\mu}}|\tilde \alpha_x\rangle = |\alpha_x\rangle - e^{-\mu/2}|0\rangle $ where $x$ can be $A, B$. We only need consider the whole-space state of $|\tilde \alpha_x\rangle$. Therefore, we need to consider
the corresponding whole-space states in the following form instead of the ideal states:
\begin{align}\label{whole}
|0\rangle&\longrightarrow |0\rangle\nonumber\\
 |\alpha_A\rangle&\longrightarrow e^{-\mu/2}|0\rangle + \sqrt{1-e^{-\mu}} |\psi_A(\tilde \alpha_A)\rangle\nonumber\\
 |\alpha_B\rangle&\longrightarrow e^{-\mu/2}|0\rangle + \sqrt{1-e^{-\mu}} |\psi_B(\tilde \alpha_B)\rangle.
\end{align}
Here the states left to the arrow are the ideal states and the states right to the arrow are the corresponding real-life states. States $|\psi_A(\tilde \alpha_A)\rangle$ and $|\psi_B(\tilde \alpha_B)\rangle$ contain whatever side-channel information such as the frequency spectrum, the polarization, the wave shape, the emission time and so on. {\em We assume Eve exactly knows all these information.} These mean, Eve knows all details of the state in the whole space and she can write down the whole-space states exactly and she can make use of it in whatever way she likes. However, consider the form of Eq.(\ref{whole}), it is obvious that there exists a unitary operation relates the ideal states and the whole-space states used in the protocol. Note that here we don't assume errors in the operational space. This means when we assume a non-ideal whole-space state, it must be able to present the same result in the operational space with that of the ideal state. Say, for state $|\psi_A(\tilde\alpha_A)\rangle$, if it is measured in photon-number space, it can only present the same probability distribution $P_n$ for
different photon number states with ideal state $|\tilde\alpha\rangle$ result, i.e., $ |\langle n|\psi_A(\tilde\alpha_A)\rangle|^2 = |\langle n|\tilde\alpha_A\rangle|^2 $. Given this, we immediately know that $\langle 0 |\psi_A(\tilde\alpha_A)\rangle=\langle 0 |\psi_B(\tilde\alpha_B)\rangle =0$
Therefore there exists the following unitary transformations:
\begin{align}\label{ut}
& \mathcal U_A |0\rangle =\mathcal U_B|0\rangle =|0\rangle\nonumber
& \mathcal U_A |\tilde\alpha_A\rangle =|\psi_A(\tilde\alpha_A)\rangle\nonumber\\
& \mathcal U_B |\tilde\alpha_B\rangle =|\psi_B(\tilde\alpha_B)\rangle
\end{align}
These equations show that the real-life source which emits non-ideal whole-space states can be mapped from an ideal source by two mode unitary transformation $\mathcal U_A\otimes \mathcal U_B$. Say, given an ideal two mode source in the protocol, one can simply take unitary transformation $\mathcal U_A$ to every encoding state from
Alice's ideal physical source and take a unitary transformation $\mathcal U_B$ to every encoding state from Bob's ideal physical source,
{\em they} can in this way obtain all  their imperfect states of the real-life source. Applying our {\em Theorem} 1, we immediately conclude that our real protocol, protocol R is secure with a real-life source if it is secure with an ideal source.

Obviously, the protocol allows to use a source with unstable side-channel information. Say, at any time window $i$, {\em they} have prepared the whole-space candidature states $|\psi_{Ai}(\tilde\alpha_A)\rangle$ and $|\psi_{Bi}(\tilde\alpha_B)\rangle$, we only need to replace the unitary transformation $\mathcal U_{A}$ and
$\mathcal U_{B}$ in Eq.(\ref{ut}) by $\mathcal U_{Ai}$ and  $\mathcal U_{Bi}$.
\\Note 5 Intensity difference does not affect the security.
Though we demonstrate the protocol with the condition $|\alpha_A|=|\alpha_B|=\mu$, however, this condition is not needed for security. We only need that the {\em intensities of all time windows are upper bounded by $\mu$ }. Say, at any individual time window $i$, we have $|\alpha_{Ai}|=\sqrt{\mu_{Ai}}\le\sqrt{\mu}$ and $|\alpha_{Bi}|=\sqrt {\mu_{Bi}}\le\sqrt{\mu}$. All these states can be obtained from an imagined coherent state of intensity $\mu$ by attenuation. Again, applying our
{\em Theorem } 1 we can assume that {\em they} are using  coherent states with stable and exact intensity $\mu$\cite{apl}.
\\Discussions: Revised protocol by post selection. In the protocol, to produce a good key rate, we need Charlie take phase compensation effectively.
Technically, we can further simplify the protocol without such type of active operation. Instead, Alice and Bob can do post selection, {\em they} only use those effective events whose corresponding initially prepared state $|\alpha_A\rangle=|\sqrt\mu e^{i\gamma_A}\rangle$, $|\alpha_B\rangle=|\sqrt\mu e^{i\gamma_B}\rangle$ in the time window satisfying
  \begin{equation}
  1-|\cos(\gamma_{B}-\gamma_{A})| \le |\lambda |.
  \end{equation}
  If we take a very small $|\lambda|$ value, the phase-flip error rate will be small, though the data size is also small.

\begin{figure}
    \includegraphics[width=250pt]{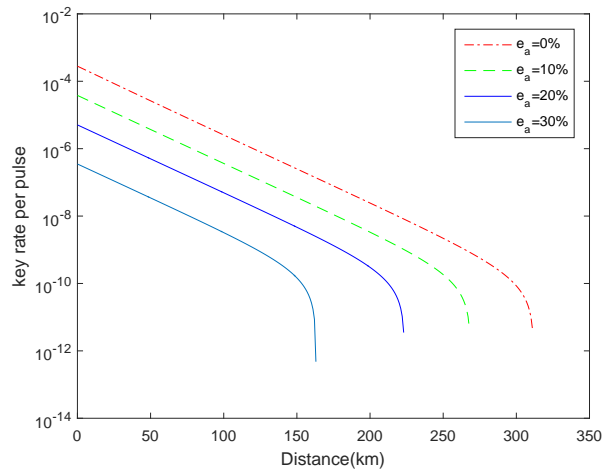}
    \caption{Log scale of the key rate as a function of the distance between Alice and Bob with different misalignment error rate. $e_a$: single-photon misalignment error. }\label{fig:result}
\end{figure}

Though the protocol presented above is source-side-channel free, it is no entirely source-independent, such as those device-independent protocols \cite{ind1} and the protocol shown in\cite{ind3}. Our protocol takes no assumption for the side channel space, but it has conditions in the operational space. The security proof in its present form assumes an exact vacuum and the sub-Poisson distribution in Fock space for the coherent state. Possibly, these conditions can be loosened through the worst-case study in the future. However we believe even in its present form it has obvious advantages in security already. \\
{\em Numerical simulation.}
Assume detector dark count rate to be $10^{-11}$ with detection efficiency of $80\%$,  a linear lossy channel  with transmittance $\eta=0.1^{-L/100km}$, and the correction efficiency is $f=1.1$. The results of numerical simulation are shown in Fig. \ref{fig:result}.
\section*{Appendix}
\subsection{Outline}
There are two parts for the security proof here. Part 1 includes virtual protocols and reduction. In part 1,  through virtual protocols and
reductions that  we first show the security if {\em they} only used $\tilde Z$-windows. In a real protocol, {\em they} sometimes use $\tilde Z$-windows and sometimes use
other time windows. We regard effective bits from $\tilde Z$-windows as un-tagged bits and the effective bits from other time windows as tagged bits. Applying the tagged model\cite{gllp} we can obtain the key rate formula as Eq.(\ref{kr00}). In the formula, we need two parameters, the bit-flip error rate and the phase-flip error rate.
A bit-flip error is the case that Alice and Bob have different values for an effective bit. Therefore the bit-flip error rate can be directly tested by the subset $v$.
The phase-flip error  is originally defined on the phase-flip rate of virtual ancillary photons. Which can be verified by observing the effective events of  $X$-windows in a virtual protocol, heralded by different detectors (left-detector, $L$, or right detector, $R$). The $X$-windows in a virtual protocol contains two subsets, an $X_+$-window that sends out a real-photon state $\rho_+=|\chi^+\rangle\langle\chi^+|$ and an $X_-$-window that sends out a real-photon state
$\rho_-=|\chi^-\rangle\langle\chi^-|$ and
\begin{equation}\label{chi}
|\chi^{\pm} \rangle= (|0,\alpha_B\rangle\pm |\alpha_A,0\rangle) / \mathcal N_{\pm}
\end{equation} where $1/\mathcal N_{\pm}$ are normalization factors for states $|\chi^{\pm}\rangle$.
In our virtual protocol, {\em they} set a probability ${|\mathcal N_+|^2}/4$ for an $X_+$-window and a probability ${|\mathcal N_-|^2}/4$ for an $X_-$-window
 whenever they use an $X$-window. The sent out states of an $X$-window is
 \begin{equation}\label{xd}
\rho_X=|\mathcal N_+|^2{\rho_+}/4+|\mathcal N_-|^2{\rho_-}/4.
\end{equation}
As shall be shown, the phase flip error rate is
 \begin{align}
 & e^{ph} = \frac{n^R_{X_+} + n^L_{X_-}}{n_X}
 =  \frac{n^R_{X_+} + n^L_X-n^L_{X_+}}{n_X}\nonumber\\
 &\le \bar e^{ph} = \frac{\bar n^R_{X_+} + n^L_X-{\underline n}^L_{X_+}}{n_X}
 \end{align}
 where $n^d_a$ is the number of effective $X$-windows heralded by joint events of $d$ and $a$, and $d=L,R$, $a=X_+,X_-,X$; and $n_X = n_X^L + n_X^R$.
Event $d$: Charlie announces that detector $d$ has clicked and the other detector has not click. Event $a$: it is a time window of $a$.
However, in a real protocol, we don't have such a state, and hence we have no way to obtain the value by direct observation. But we can still verify the upper bound of the phase-flip error rate by observing data of other states in the real protocol with calculation formulas presented in Part 2.
Say, in the real protocol, {\em they} randomly take two subsets of all time windows, $v$ and $u$. Details on {\em sending} or {\em not-sending} of $u$ are never
announced. But we know that set $u$ contains a number of $\tilde Z$ windows whose density operator is
\begin{equation}
\rho_{\tilde Z} = \frac{1}{2}(|0,\alpha_B\rangle\langle|0,\alpha_B|+|\alpha_A,0\rangle\langle\alpha_A,0|)
\end{equation}
 Obviously,
the density operator has another equivalent convex and
\begin{equation}
 \rho_{\tilde Z} = \rho_{X}
\end{equation}
This means, in a virtual protocol, if we don't announce any information on {\em sending} or {\em not-sending} of time windows in set $u$, to Eve, those $\tilde Z$-windows in set $u$ are identical to $X$-windows. Say, in a real protocol, set $u$ contains fake $X$-windows which are identical to true $X$-windows to Eve.
We can then verify the value of $\bar e^{ph}$ of fake $X$-windows by observing and calculating the data of another subset of time windows, set $v$ in the real protocol.
  Explicitly, as shall be shown in part 2, we need the following data in the calculation:
\\1, The $d$-event rate   of time windows $A$, denoted by $S^d_A$.
 It is just the fraction of effective time windows heralded by event $d$ from $A$-windows. Here $d=L,R$, for an effective event of detector $d$ clicking and the other detector not clicking as announced by Charlie. We need the values of  $S^d_A$ for $d=L,R$ and $A=\tilde Z, \mathcal B, \mathcal O$, where $\mathcal B$ is for a time window when both Alice and Bob decide {\em sending}, and $\mathcal O$ is a time window when both Alice and Bob decide {\em not-sending}.
  As shall be shown in Part 2, based on these observed data, we can then calculate the upper bound of phase-flip error rate, $\bar e^{ph}$. Explicitly
 \begin{equation}\label{phasef}
    e^{ph}\le \overline{e}^{ph} = \frac{(1+e^{-2\mu}) \left[\overline{S}_{X_+}^{R}-\underline{S}_{X_+}^{L}\right] +2S_{\tilde Z}^{L}} {2(S_{\tilde Z}^L +S_{\tilde Z}^R)}
\end{equation}
where  $\overline{S}_{X_+}^{d}$ ($ \underline{S}_{X_+}^{d}$) is the upper bound (lower bound) of
${S}_{X_+}^{d}$, with $d=L,R$ and
\begin{equation}\label{up}
\begin{split}
    \overline{S}_{X_+}^{d} =& \frac{1}{2(1+e^{-\mu})} \{e^{-\mu} S_{\mathcal O}^{d} + \frac{1}{e^{-\mu}} S_{\mathcal B}^{d} + \frac{(1-e^{-\mu})^2}{e^{-\mu}} \\
    &+  2\sqrt{S_{\mathcal O}^{d} S_{\mathcal B}^{d}} + 2(1-e^{-\mu}) \sqrt{S_{\mathcal O}^{d}}
    \\& + \frac{2(1-e^{-\mu})}{e^{-\mu}} \sqrt{S_{\mathcal B}^{d}} \}
    \ge  S_{X_+}^{d}
\end{split}
\end{equation}
\begin{equation}\label{down}
\begin{split}
     \underline{S}_{X_+}^{d} =& \frac{1}{2(1+e^{-\mu})} \{e^{-\mu} S_{\mathcal O}^{d} + \frac{1}{e^{-\mu}} S_{\mathcal B}^{d}\\&- [2\sqrt{S_{\mathcal O}^{d} S_{\mathcal B}^{d}} + 2(1-e^{-\mu}) \sqrt{S_{\mathcal O}^{d}}
    \\& + \frac{2(1-e^{-\mu})}{e^{-\mu}} \sqrt{S_{\mathcal B}^{d}}] \} \le S_{X_+}^{d}
\end{split}
\end{equation}
\subsection{Part 1, virtual protocols, reduction, and key rate from tagged model}
 \noindent
 {\bf Definitions} of {\em effective event}: We define an {\bf effective event}  if Charlie announces one and only one detector clicking for an individual  $Z$-window.  {\em They} will then only use states or data corresponding to  effective events in the protocol. A time window that presents an effective event is named as an effective time window. An {\em effective ancillary photon} is an ancillary
  photon corresponding to an effective event. A classical bit from an effective time window is named as an effective bit.
  \\Event $L$ or $L$-event: an effective event of the left detector clicking and the right detector  not-clicking;
  Event $R$ or $R$-event: an effective event of the right detector clicking and the left detector  not-clicking.
\subsubsection{Virtual protocol V1}
\noindent{\em Preparation stage}\\
 {\em They} pre-share classical information for different time windows {\em they} will use, $X$-windows and $Z$-windows. {\em They} also pre-share
  an extended state
\begin{align}\label{omp}
&\Omega_{i} = |\Psi_{i}\rangle\langle\Psi_{i}|\nonumber\\
& |\Psi_{i}\rangle = \frac{1}{\sqrt 2}(|0,\alpha_B\rangle\otimes|01\rangle+|\alpha_A,0\rangle\otimes|10\rangle)
\end{align}
for the $i$th time window.
Here $|\alpha_A|=|\alpha_B|=\sqrt\mu$, $|\alpha_{Ai}\rangle=|\sqrt\mu e^{i\gamma_{Ai}}\rangle $, $|\alpha_{Bi}\rangle=|\sqrt\mu e^{i\gamma_{Bi}}\rangle $.
{\em They} announce the states including the global phases $\gamma_{Ai},\gamma_{Bi}$

For presentation simplicity, we shall omit the subscripts $i$ in all phase values $\gamma_{A_i},\gamma_{B_i}$ and states.
Also, we introduce states $|\chi^+\rangle, |\chi^-\rangle$ as defined by Eq.(\ref{chi}) for the extended state. Explicitly, it can be written in
\begin{equation}
|\Psi\rangle =\left(\mathcal N_+|\chi^+\rangle\otimes|\Phi^0\rangle + \mathcal N_-|\chi^-\rangle\otimes |\Phi^1\rangle\right)/ 2
\end{equation}
and $|\Phi^0\rangle=\frac{1}{\sqrt 2}(|01\rangle+|10\rangle)$, $|\Phi^1\rangle=\frac{1}{\sqrt 2}(|01\rangle-|10\rangle)$.\\
{\em Virtual Protocol} V1\\
V1-1 At any time window $i$, no matter it is a $Z$-window or an $X$-window, {\em they} send out to Charlie the real-photon state from state $\Omega$  as defined
by Eq.(\ref{omp})
 and keep the ancillary photons locally.
\\V1-2 Charlie announces his measurement outcome of  all time windows.  This announcement determines the effective time windows.
    \\ \underline{ Definition}: {\em They} can now divide {\em their} time windows into 4 subsets, $X^L,X^R,Z^L,Z^R$ where a time window of $\mathcal W^d$ is an effective
    $\mathcal W$-window heralded by detector $d$ clicking and the other detector not clicking. $\mathcal W = X,Z$, and $d=L,R$. Also, we shall use $\mathcal A_{\mathcal W^d}$ for the set of effective ancillary photons of time window $\mathcal W^d$.
 \\V1-3  {\em They} check the phase-flip error rate $e^{ph}$ for set of $\mathcal A_{ X^d}$, where $d=L,R$, which is also the estimated phase-flip error rates of set $\mathcal A_{ Z^d}$.
 \\V1-4 {\em They }  purify the  ancillary photons of sets $ \mathcal{A}_{Z^L}$ and $\mathcal A_{Z^R}$  separately. After purification, {\em they} obtain high quality single-photon states $|\Phi^0\rangle$ or $|\Phi^1\rangle$ with (almost) 100\% purity.  {\em They} each measures the photon number locally to the purified photons and obtain the final key $k_f$. Alice puts down a bit value 0 or 1 whenever she obtains a measurement outcome of vacuum or 1 photon, Bob puts down a bit value 1 or 0 whenever she obtains a measurement outcome of vacuum or 1 photon.
\\ \underline {Note 1} {\em Security }. The security of the final key is based on the faithfulness of the purification\cite{t2}, i.e., the estimation of phase-flip error rate. Charlie has determined  effective ancillary photons but Alice and Bob test the phase-flip error rate themselves in step 1-3.
 The extended state of an $X$-window  identical to that of a $Z$-window, therefore  the phase-flip-error rate value of set $\mathcal A_{ X^d}$ is exactly the  value of set $\mathcal A_{ Z^d}$.
\\ \underline{ Note 2} { Definitions of phase-flip-error rate}. \\
  Suppose set $\mathcal A_{X^d}$ contains $n^d$ effective ancillary photons. If each photon of set $\mathcal A_{X^d}$ was measured in basis $\{|\Phi^0\rangle, |\Phi^1\rangle\}$ and there were
   $n_0^d$ outcome of $|\Phi^0\rangle\langle\Phi^0|$, and $n_1^d$ outcome of $|\Phi^1\rangle\langle\Phi^1|$, the phase-flip error rate for set $\mathcal A_{ X^d}$ is
   \begin{equation}\label{er}
   e^{ph}= \frac{{ \min}\left(n_0^d,n_{1}^d\right)}{n^d}.
   \end{equation}
   Changing the values of $n_0^d,n_1^d,n^d$ into the corresponding values of set $\mathcal A_{Z^d}$ in Eq.(\ref{er}), we can define the phase flip error rate for set $\mathcal A_{Z^d}$. Statistically,  $e^{ph}$ for set
   $\mathcal A_{X^d}$  is also
    the asymptotic phase-flip error rate of set $\mathcal A_{ Z^d}$.
    To know
   the values $e^{ph}$, {\em they} can choose to measure each photon of set $\mathcal A_{X^d}$ in basis $\{|\Phi^0\rangle, |\Phi^1\rangle\}$. But instead of this, {\em they} can also choose to take  local measurements in basis $\{|x\pm\rangle\}$ in each side and check the parity of each  measurement outcome. (Outcome of $|x+\rangle|x+\rangle$ or
    $|x-\rangle|x-\rangle$) are even-parity while $|x+\rangle|x-\rangle$ or $|x-\rangle|x+\rangle$ are odd parity.) Note that all effective ancillary photons are single-photons. As it is easy to see , for single-photons, the fraction of odd parity (even parity) outcome from  measurement of each side in basis $\{|x\pm\rangle\}$
   is exactly equal to the fraction of $|\Phi^1\rangle\langle\Phi^1|$ ($|\Phi^0\rangle\langle\Phi^0|$) outcome from the measurement in basis  $\{|\Phi^0\rangle, |\Phi^1\rangle\}$. Moreover, this measurement step is only needed here for this Virtual protocol, it is not needed for a real protocol. For ease of presentation,
   we suppose {\em they} use the measurement basis $\{|\Phi^0\rangle,|\Phi^1\rangle\}$.
  \\ \underline{Note 3} { Reduction of pre-shared states for $X$-windows}\\
   \underline{{\em Reduction 1}}
  It makes no difference to anyone outside  if {\em they} measure all ancillary photons of $X$-windows in basis $\{|\Phi^0\rangle,|\Phi^1\rangle\}$
   before the protocol starts.  After measurement to the ancillary photon, {\em they} obtain one of the following outcome extended state  for an $X$-window, depending on the measurement outcome of ancillary photon:\\
   either
   \begin{equation}\label{s0}
    |\chi^+\rangle\langle\chi^+|\otimes |\Phi^0\rangle\langle\Phi^0|
   \end{equation}
   with probability $|\mathcal N_+|^2/4$ and
   \begin{equation}\label{s1}
   |\chi^-\rangle\langle\chi^-|\otimes |\Phi^1\rangle\langle\Phi^1|
    \end{equation} with probability $|\mathcal N_-|^2/4$.
    \\ \underline{{\em Reduction 2}}
    Alternatively, {\em they} can just start with states of Eq.(\ref{s0},\ref{s1}) for their $X$-windows. {\em They} need pre-share classical information on $Z$-windows, $X_+$-windows, and $X_-$-windows. The pre-shared classical information for $X$-windows assign a probability $|\mathcal N_+|^2/4$ for $X_+$-window and a probability $|\mathcal N_-|^2/4$ for $X_-$-window.
    {\em They}  pre-share real-photon states $| \chi^+\rangle\langle\chi^+|$ for $X_+$-windows and $|\chi^-\rangle\langle\chi^-|$ for $X_-$-windows. Imagine that {\em they} also pre-share some single-photon states $|\Phi^0\rangle$ and $|\Phi^1\rangle$. (These states $|\Phi^0\rangle$ and $|\Phi^1\rangle$ are not really necessary, to show everything clearly we assume so at this moment.)
\\In an $X_+$-window,
  {\em they} label a pre-shared state $|\Phi^0\rangle$ as the ancillary photon for this state $|\chi^+\rangle$ above and have an extended state
  \begin{equation}\label{kaka1}
  \Omega_{+} = |\chi^+\rangle\langle \chi^+| \otimes |\Phi^0\rangle\langle\Phi^0|
  \end{equation}
  {\em They} then send the real-photon state $|\chi^+\rangle$  out to Charlie.
  In an $X_-$-window, {\em they} label a pre-shared state $|\Phi^1\rangle$ as the ancillary photon for this state $|\chi^-\rangle$ above and have an extended state
  \begin{equation}\label{kaka2}
  \Omega_{-} = |\chi^-\rangle\langle \chi^-| \otimes |\Phi^1\rangle\langle\Phi^1|
  \end{equation}
  {\em They} then send the real-photon state $|\chi^-\rangle$  out to Charlie.
  Here we have used the same definition for $|\chi^+\rangle,|\chi^-\rangle$ as  Eq.(\ref{chi}).
 \\
 From Eq.(\ref{kaka1}) and Eq.(\ref{kaka2}) we can see that, in an $X_+$-window, the ancillary-photon state must be $|\Phi^0\rangle$; in an $X_-$-window, the ancillary-photon state must be $|\Phi^1\rangle$.
     \\
     Therefore, according to our {\em  Definition  1},  {\em they} can use  the following more operable definition to calculate each quantities in Eq.(\ref{er})
     \begin{align}
     & n_0^d = n_{X_+}^d \label{er1}\\
     & n_1^d = n_{X_-}^d\label{er2}
     \end{align}
     for Eq.(\ref{er}). Here $n_{X_{\pm}}^d$ is the number of $X_+$-windows or $X_-$-windows heralded by detector $d$ clicking and the other detector not clicking,  and $d=L,R$,
     $L$ for left detector and $R$ for right detector.
     \begin{equation}\label{er3}
   e^{ph}= \frac{{ \min}\left(n_{X_+}^d,n_{X_-}^d\right)}{n^d}.
   \end{equation}
     Given this phase-flip error rate formula,  the ancillary photons for $X$-windows  are actually {\em not} needed in the protocol.
     \\ \underline{Note 4} {\em quasi-purification}  \\Since {\em their} goal is to have the final key only, a true purification to ancillary photons is not necessary\cite{simple}.
{\em They} can choose to    measure all ancillary photons of $Z$-windows in advance\cite{simple} in photon-number basis and then take virtual purification to classical data of $Z$-windows corresponding to those effective events.  {\em They} then take a virtual quasi-purification to the classical data, which is just the final key distillation. Also, the pre-shared extended state for a $Z$-window is just
 $(|0,\alpha_B\rangle\langle 0,\alpha_B| \otimes |01\rangle\langle 01| + |\alpha_A,0\rangle\langle \alpha_A,0| \otimes |10\rangle\langle 10|)/2$.
 \\ \underline{Note 5} Purifying all effective ancillary photons in one batch. Definitely, {\em they} can choose to purify all effective
 ancillary photons of $Z$-windows in one batch. The phase-flip error rate is
 \begin{equation}\label{phat}
  e^{ph} = \frac{\min\left(n^L_{X_+},n^L_{X_-}\right)+\min\left(n^R_{X_+},n^R_{X_-}\right) }{n_X}
 \end{equation}
 where $n_X=n^L_{X_+}+n^L_{X_-}+n^R_{X_+}+n^R_{X_-}$ is the number of all effective $X$-windows.
 Surely, $n_{X_-}^L\ge \min\left(n^L_{X_+},n^L_{X_-}\right)$ and $n_{X_+}^R\ge \min\left(n_{X_+}^R , n_{X_-}^R\right)$.
  Therefore the phase-flip error rate formula of Eq.(\ref{phat}) can be simplified into
 \begin{equation}\label{phases2}
 e^{ph} \le \frac{n_{X_-}^L+n_{X_+}^R}{n_X}
 \end{equation}
 which is simply first to count the number of effective $X_-$-windows heralded by left detector and also the effective $X_+$-windows heralded by the right detector and then take the rate of errors per effective $X$-window.
 \\ If {\em they} use this formula, Charlie can make a high quality raw state of effective ancillary photons for Alice and Bob by setting his measurement set-up
 properly so that
 with very small probability for the left-detector-clicking (right-detector-clicking) due to the incident state of $|\chi^-\rangle$ ($|\chi^+\rangle$).
\subsubsection{Virtual protocol V2}
\noindent {\em Preparation stage} \\
Here we assume {\em they} pre-share a mixed state of
\begin{equation}
\Omega=\sum_y p_y \Omega_y
\end{equation}
 $y=0,1,\mathcal B,\mathcal O,+,-$, $p_0=p_1$ and
 \begin{align}
 \Omega_0=|0,\alpha_B\rangle\langle 0,\alpha_B|\otimes |01\rangle\langle 01|\\
 \Omega_1=|\alpha_A,0\rangle\langle \alpha_A,0|\otimes |10\rangle\langle 10|\\
 \Omega_{\mathcal B}=|\alpha_A,\alpha_B\rangle\langle \alpha_A,\alpha_B|\otimes |11\rangle\langle 11|\\
 \Omega_{\mathcal O}=|0,0\rangle\langle 0,0|\otimes |00\rangle\langle 00|\\
 \Omega_+=|\chi^+\rangle\langle \chi^+| \otimes |2,2\rangle\langle 2,2|\\
 \Omega_-=|\chi^-\rangle\langle \chi^-| \otimes |3,3\rangle\langle 3,3|.
 \end{align}
{\em They} don't pre-share any classical information for time windows. For clarity in presentation, we  define different kinds of time windows
by the local ancillary states:
$|01\rangle\langle 01|$ for $Z_0$, $|10\rangle\langle 10|$ for $Z_1$, $|11\rangle\langle 11|$ for $Z_{\mathcal B}$, $|00\rangle\langle 00|$ for  $Z_{\mathcal O}$,
$|2,2\rangle\langle 2,2|$ for  $X_+$, and $|3,3\rangle\langle 3,3|$ for  $X_-$.
\\ Time window $Z_0$ ($Z_1$) corresponds to the case Alice (Bob) decides {\em not-sending} and Bob (Alice) decides {\em sending}. Time window $Z_{\mathcal B}$
 ($Z_{\mathcal O}$) corresponds to the case that both Alice and Bob decide {\em sending} ({\em not-sending}).
\\
   {\em They} don't know which time window belongs to which kind of windows say
   $Z_y$-window for $y=0,1,\mathcal{B},\mathcal{O}$ or $X_y$-window for $y=+$ or $-$.
   But {\em they} know how much $\tilde Z$-windows ($Z_0$-windows or $Z_1$-windows) there and also by announcing details of a random subset of $Z$-windows, {\em they} can judge the fraction of effective time windows among all $\tilde Z$-windows thus {\em they} know the total number of effective ${\tilde Z}$-windows among all effective $Z$-windows. Then {\em they} can apply the tagged mode to calculate the final key.
\\{\em Virtual Protocol} V2\\
V2-1 At any time window $i$, {\em they} send out to Charlie the real-photon state
 from pre-shared extended state $\Omega$, $b=0,1,{\mathcal B},{\mathcal O}$. By measuring the local ancillary-photon, {\em they} each know whether the time window is an $X$-window or an $Z$-window. Explicitly, a local state $|2\rangle\langle 2|$ for an $X_+$-window and a local state $|3\rangle\langle 3|$ for an $X_-$-window. A local state $|0\rangle\langle 0|$ or $|1\rangle\langle 1|$
for a $Z$-window. Though {\em they} know which time windows are $Z$-windows, {\em they} don't know the information on which kinds of $Z$-windows, ($Z_0,Z_1,Z_{\mathcal B}, Z_{\mathcal O}$).
 \\V2-2 Charlie announces his measurement outcome of  all time windows.
    \\V2-3  According to Charlie's announcement, {\em they} are aware of those effective $Z$-windows and $X$-windows.
For an $Z$-window, Alice (Bob) puts down a classical bit 0 (1) if her (his) local ancillary state is vacuum and puts down a classical bit 1 (0) if her (his) local ancillary state is one-photon.  {\em They} randomly take a subset $v$ of $Z$-windows, announcing local measurement outcome of each time windows in set $v$ so that {\em they} can judge the asymptotic value of bit-flip error rate $E_Z$ by counting the number of effective $\tilde Z$-windows from all effective $Z$-windows of set $v$.
    {\em They} estimate the phase-flip error rate $e^{ph}$ effective bits of $\tilde Z$-windows (time windows of $Z_0$ and $Z_1$) by Eq.(\ref{phases2}) through observing events of all $X$-windows.
 \\V2-4 {\em They} regard the effective bits from $\tilde Z$-windows as un-tagged bits and the effective bits from time windows of $Z_{\mathcal B}$ and $Z_{\mathcal O}$ as tagged bits. Applying the tagged model\cite{gllp}, {\em they }   distill effective bits of $Z$-windows and obtain the final key $k_f$ with the length given by Eq.(\ref{kr00}).
 \\ \underline{Note 1} A bit-flip error is an effective $Z$-window when Bob's bit value is different from Alice's. {\em They } can  verify it by testing a random subset of $Z$-windows.
 \\ \underline{Note 2}  Estimating  $\bar e^{ph}$ the upper bound of phase flip error rate by observing subset $v$ of $Z$-windows only.
 Consider Eq.(\ref{phases2}). It is equivalent to
 \begin{equation}
 e^{ph} \le \frac{n_{X_-}^L+n_{X_+}^R}{n_X}\le \bar e^{ph} =\frac{\bar n_{X_+}^R-\underline{n}_{X_+}^L+n_X^L}{n_X}
 \end{equation}
 here bar represent upper bound and underline represent lower bound. $n_X$: total effective $X$-windows, $n_X^L$: number of those effective $X$-windows heralded by detector $L$. $n_a^d$: number of effective $a$-windows heralded by detector $d$, $a=X_+,X_-$ and $d=L,R$.
    \\ \underline {Note 4} The $X$-windows are not really needed. Because {\em they} can obtain the bound values $\bar n_{X_+}^R$ and $\underline{n}_{X_+}^L$ by observing a subset of $Z$-windows, instead of observing $X$-windows. Therefore we can use fake $X$-windows to replace the original $X$-windows, provided that the real-photon state of the fake $X$-windows is identical to that of the original $X$-windows. Obviously, in the virtual protocol V2, a $\tilde Z$-window is a perfect fake $X$-window.
  \subsubsection{Virtual protocol V3}
  \noindent {\em Preparation stage} \\
  {\em They} pre-share an extended  mixed state of
  \begin{equation}\label{om5}
  \Omega = \sum_y p_y \Omega_y
  \end{equation}
  and $p_0=p_1$.
  \begin{align}
  & \Omega_0=|0,\alpha_B\rangle\langle 0,\alpha_B|\otimes|01\rangle\langle 01|\nonumber\\
  & \Omega_1= |\alpha_A,0\rangle\langle \alpha_A,0|\otimes |10\rangle\langle10|\nonumber\\
  & \Omega_{\mathcal B} = |\alpha_A,\alpha_B\rangle\langle \alpha_A,\alpha_B|\otimes |11\rangle\langle11|\nonumber\\
  &\Omega_{\mathcal O}=|0,0\rangle\langle 0,0|\otimes |00\rangle\langle00|.
  \end{align}
  \\In this virtual protocol, {\em they} don't pre-share any classical information, but we can still {\em define} time windows $Z_0,Z_1,Z_{\mathcal B},Z_{\mathcal O}$
  by the four different ancillary states. Say, if {\em they} each announce the local measurement outcome of ancillary-photon state at a certain time window $i$, {\em they} can know the specific kind of time window for $i$. In the protocol {\em they} will take a random subset of time windows $v$ for error test. {\em They} each announce  local measurement
  outcome  of the ancillary state of every time window in set $v$.
\\{\em Virtual Protocol} V3\\
V3-1 At any time window $i$, {\em they} send out their real-photon state.
 \\V3-2 Charlie announces his measurement outcome of  all time windows.
    \\V3-3  Alice (Bob) measures her (his) ancillary photon for all effective time windows, outcome vacuum for a classical bit 0 (1) and outcome 1-photon for a bit value
    1 (0). Through classical communications, {\em they}  take two random subsets $v$, $u$ for error test. {\em They} don't announce information of  random subset $u$.  Set $u$ contains a subset of $\tilde Z$-windows whose real-photon states sent out make a perfect fake state for the imaginary $X$-windows as defined in the Virtual protocol V2. To every time window of set $v$, {\em they} each announce the local measurement outcome of ancillary-photon state.
     From this announced outcome, {\em they} can know the bit-flip error rate $E_Z$, and the fraction of un-tagged bits, i.e., bits from time windows $Z_0$ or $Z_1$.
     {\em They} also know  the yield values of $S^d_y=\frac{n_y^d}{N_y}$ where $n_y^d$ is the number of effective windows heralded by event $d$ from all $Z_y$-windows,
     $y=\mathcal B, \mathcal O$, i.e., the rate of effective windows heralded
     by detector $d$ for all time windows $Z_y$, where $d=L,R$ and $y=\mathcal B, \mathcal O$.  Using these values {\em they} calculate  the upper bound value of
     $\bar e^{ph}$ of Eq.(\ref{phases2}) for the  fake $X$-windows ($\tilde Z$-windows of set $u$).
 \\V3-4 {\em They }   distill effective bits of $Z$-windows and obtain the final key $k_f$. Applying the tagged model\cite{gllp}, {\em they} have the  length of final key by Eq.(\ref{kr00}).
 \\ \underline{Note 1} {\em They} can produce the state of Eq.(\ref{om5}) locally. Setting $p_0=p_1=q(1-q)$, $p_{\mathcal O}= (1-q)^2$ and $p_{\mathcal B}=q^2$, {\em they} can produce the state $\Omega$ of Eq.(\ref{om5}) locally in this way: At every time $i$, Alice (Bob) randomly decides {\em sending} with probability $q$ or {\em not-sending} with probability $(1-q)$. For a decision {\em sending}, she (he) sends out a coherent state $|\alpha_A\rangle\langle\alpha_A|$ ($|\alpha_B\rangle\langle\alpha_B|$) to Charlie, puts down a bit value 1 (0), and produces a local state $|1\rangle\langle 1|$. For a {\em not-sending} decision,  she (he) sends out a vacuum to Charlie, puts down a bit value 0 (1), and produces a local state $|0\rangle\langle 0|$. If {\em they} produce the state $\Omega$ of Eq.(\ref{om5}) in this way, {\em they} don't need to pre-share anything. Also,  without the local ancillary-photon state {\em they} can still complete the final key distillation therefore the local ancillary state is actually not needed. This comes back to the real protocol.
\subsection{Part 2, Phase-flip error rate}
\subsubsection{ Input-Output model}
 Consider an input state to Charlie sent from Alice (and Bob). Alice will observe Charlie's instrument $\mathcal L$ for the corresponding outcome (classical outcome). Charlie has no access to Alice's source.

Suppose in the beginning of a certain time window, Charlie receives a state $\ket\psi$. We shall call this as input state to Charlie. Consider the extended state
made up of the input and Charlie's state $|\kappa\rangle$. Charlie's  instrument state $\mathcal L$ is included in the ancillary state $|\kappa\rangle$. The initial state is
  \begin{equation}|\Psi_{ini}\rangle = |\psi\rangle\otimes|\kappa\rangle.\end{equation}
  At time $t$ Charlie observes his instrument $L$ to see the result. His instrument $\mathcal L$ is observed by Alice and she can find  result from $\{l_i\}$ accompanied with its eigenstate $| l_i\rangle$ then.
  Most generally, after state $|\psi\rangle$ is sent to Charlie, Charlie's initial state $|\Psi_{ini}\rangle = |\psi\rangle\otimes|\kappa\rangle $ will evolve with time under a quantum process. Here we  assume a unitary quantum process $\mathcal U$.  Even though Charlie presents a non-unitary quantum process, it can be represented
  by a unitary process through adding more ancillary states. So, given the general ancillary state $|\kappa\rangle$, we can simply assume a unitary quantum process for Charlie.  At time $t$, the state is now
 \begin{equation}
 |\Psi(t)\rangle = \mathcal U(t)|\Psi_{ini}\rangle=\mathcal U(t)(|\psi\rangle\otimes|\kappa\rangle)
 \end{equation}
 In general, the state at time $t$ can be written in a bipartite form of another two subspaces, one is the instrument space $\mathcal L$ and the other is the remaining part of the space, subspace $\bar {\mathcal L}$.
 Given the initial input state $|\psi\rangle$ to Charlie, the probability that he observes the result $l_1$ at time $t$ is
 \begin{equation}\label{p0}
 p^{l_1}=\langle l_1|\tr_{\bar{\mathcal L}}\left(|\Psi(t)\rangle\langle\Psi(t)|\right)|l_1\rangle
 \end{equation}
We will omit $(t)$ in the following formulas.
 Suppose the space $\bar {\mathcal L}$ is spanned by basis states $\{g_k\}$, we can rewrite Eq.(\ref{p0}) by
 \begin{align}\label{basic}
 & p^{l_1} =\sum_k |\langle \gamma_k^{(l_1)}|\Psi\rangle|^2
 \end{align}
 where $|\gamma_k^{(l_1)}\rangle=|g_k\rangle|l_1\rangle$.

Since in each time window $i$ Charlie may use different quantum process with different ancillary states and different measurement, the quantum process should be written as $\mathcal U_i$ and Eq.(\ref{basic}) should be written as:
\begin{align}\label{basic_i}
    & n^{l_1} =\sum_{i=1}^K\sum_k |\langle \gamma_{i,k}^{(l_1)}|\Psi_i\rangle|^2
\end{align}
where $n^{l_1}$ is the number of time windows that Alice observes  the outcome $l_1$ from instrument $\mathcal L$ and $K$ is the total number of time windows.

Eqs. (\ref{basic})(\ref{basic_i}) are our elementary formulas for the input-output model.
\subsubsection{Calculate the outcome of one input state by observing the outcome for other states.}
Imagine different sources, source 1 emits state $\ket{\phi}$, source 2 emits state state $\ket{\phi_0}$, and source 3 emits state $\ket{\phi_1}$. State $|\phi\rangle$ has the form of
\begin{equation}\label{equ:phi}
    \ket{\phi} = c_0 \ket{\phi_0} + c_1 \ket{\phi_1} + c_2 \ket{\phi_2}
\end{equation}

 At any time window $i$, Alice only uses one source. She choose one source from $1,2,3$ randomly  with probabilities $d_1,d_2,d_3$, and, to distinguish states from different sources, she produces a local ancillary state $|1\rangle\langle1|$, $|2\rangle\langle2|$, and $|3\rangle\langle 3|$   respectively when she uses sources 1,2,3. At one time window, she will only use one source. A virtual case that Alice and Bob prepare a state with density matrix
\begin{equation}\label{equ:virtual}
\begin{split}
    \rho =& d_1\ket{\phi}\langle{\phi}| \otimes \ket1\bra1 + d_2\ket{\phi_0}\bra{\phi_0} \otimes \ket2\bra2\\
     &+ d_3\ket{\phi_1}\bra{\phi_1} \otimes \ket3\bra3,\ d_1+d_2+d_3=1
\end{split}
\end{equation}
where $\ket{k},k=1,2,3$ is the local auxiliary state.  Then they keep the auxiliary state and send out the real-photon state to Charlie. (We name the state emitted
by source 1,2, or 3 as the real-photon state.)
With the elementary formula Eq.(\ref{basic_i}) in Sec.1 and Eq.(\ref{equ:virtual}), we have the following extended state after Charlie completes operations to the real-photon state from Alice and Charlie's ancillary state:
\begin{equation}\label{equ:inter}
\begin{split}
    \tilde{\rho} =& d_1\ket{\Phi_i}\langle{\Phi_i}| \otimes \ket1\bra1 + d_2\ket{\Phi_{0i}}\bra{\Phi_{0i}} \otimes \ket2\bra2\\
     &+ d_3\ket{\Phi_{1i}}\bra{\Phi_{1i}} \otimes \ket3\bra3,\ d_1+d_2+d_3=1
\end{split}
\end{equation}
where $\ket{\Phi_i} = \mathcal{U}_i(t) (\ket{\phi}\otimes\ket{\kappa_i})$, $\ket{\Phi_{ki}} = \mathcal{U}_i (\ket{\phi_k}\otimes\ket{\kappa_i})$, $k=0,1$. On the other hand, given
Eq.(\ref{equ:phi}), we have
\begin{equation}\label{equ:Phi}
    \ket{\Phi_i} = c_0 \ket{\Phi_{0i}} + c_1 \ket{\Phi_{1i}} + c_2 \ket{\Phi_{2i}}.
\end{equation}
We can now write the formula for the number of the $l_1$-event of source $k$, $n_{k}^{l_1}$,  which is the number of time windows heralded by joint events of outcome $l_1$ for instrument $\mathcal L$ from source $k$, and $k=1,2,3$. Since we have already labeled each source by ancillary-photon states, the number of the $l_1$-event of source $k$ is just the number of joint events of outcome of $l_1$ from the instrument and outcome $|k\rangle\langle k|$ from the  measurement to the local ancillary-photon state.
\begin{equation}\label{nn1}
    n_{1}^{l_1} = d_1 \sum_{i=1}^N\sum_k |\iprod{\gamma_{i,k}^{(l_1)}}{\Phi_i}|^2
\end{equation}
\begin{equation}\label{nn2}
    n_{2}^{l_1}  = d_2 \sum_{i=1}^N\sum_k |\iprod{\gamma_{i,k}^{(l_1)}}{\Phi_{0i}}|^2
\end{equation}
\begin{equation}\label{nn3}
    n_{3}^{l_1}  = d_3 \sum_{i=1}^N\sum_k |\iprod{\gamma_{i,k}^{(l_1)}}{\Phi_{1i}}|^2
\end{equation}

Therefore, we have the following formula for the number of $l_1$-event from source 1:
\begin{widetext}
\begin{equation}\label{equ:nPhi}
\begin{split}
    n_{1}^{l_1}/d_1=\sum_{i=1}^N \sum_k |\iprod{\gamma_{i,k}^{(l_1)}}{\Phi_i}|^2
    =& |c_0|^2 \sum_{i=1}^N \sum_k |\iprod{\gamma_{i,k}^{(l_1)}}{\Phi_{0i}}|^2 + |c_1|^2 \sum_{i=1}^N \sum_k |\iprod{\gamma_{i,k}^{(l_1)}}{\Phi_{1i}}|^2 + |c_2|^2 \sum_{i=1}^N \sum_k |\iprod{\gamma_{i,k}^{(l_1)}}{\Phi_{2i}}|^2   \\
    &+ 2\sum_{i=1}^N \sum_k \text{Re}(c_0 c_1\iprod{\Phi_{0i}}{\gamma_{i,k}^{(l_1)}} \iprod{\gamma_{i,k}^{(l_1)}}{\Phi_{1i}}) + 2\sum_{i=1}^N \sum_k \text{Re}(c_0 c_2\iprod{\Phi_{0i}}{\gamma_{i,k}^{(l_1)}} \iprod{\gamma_{i,k}^{(l_1)}}{\Phi_{2i}}) \\
     &+ 2\sum_{i=1}^N \sum_k \text{Re}(c_1 c_2\iprod{\Phi_{1i}}{\gamma_{i,k}^{(l_1)}} \iprod{\gamma_{i,k}^{(l_1)}}{\Phi_{2i}})
\end{split}
\end{equation}
\end{widetext}
For the term $\sum_{i=1}^N \sum_k \text{Re}(c c'\iprod{\psi}{\gamma_{i,k}^{(l_1)}} \iprod{\gamma_{i,k}^{(l_1)}}{\psi'})$, with any states $\ket\psi$ and $\ket{\psi'}$, we can use Cauchy inequality
\begin{equation}\label{equ:cauchyreal}
    (\sum_{k=1}^{m} a_k  b_k)^2 \le \sum_{k=1}^{m} a_k^2 \sum_{k=1}^{m} b_k^2,\ a_k,b_k\in\mathbb{R}
\end{equation}
to obtain its bound:
\begin{equation}\label{equ:up}
\begin{split}
    &|\sum_{i=1}^N \sum_k \text{Re}(c c'\iprod{\psi}{\gamma_{i,k}^{(l_1)}} \iprod{\gamma_{i,k}^{(l_1)}}{\psi'})| \\
    \le& |c c'|\sum_{i=1}^N \sum_k |\iprod{\psi}{\gamma_{i,k}^{(l_1)}}| |\iprod{\gamma_{i,k}^{(l_1)}}{\psi'}| \\
     \le& |c c'|\sqrt{\sum_{i=1}^N \sum_k |\iprod{\psi}{\gamma_{i,k}^{(l_1)}}|^2} \sqrt{ \sum_{i=1}^N \sum_k |\iprod{\gamma_{i,k}^{(l_1)}}{\psi'}|^2}
\end{split}
\end{equation}

Recall Eq.(\ref{nn1},\ref{nn2},\ref{nn3}) we have
\begin{equation}
\begin{split}
    n_{1}^{l_1}/d_1 = \sum_{i=1}^N\sum_k |\iprod{\gamma_{i,k}^{(l_1)}}{\Phi_i}|^2, \\ n_{2}^{l_1}/d_2 = \sum_{i=1}^N\sum_k |\iprod{\gamma_{i,k}^{(l_1)}}{\Phi_{0i}}|^2, \\
    n_{3}^{l_1}/d_3 = \sum_{i=1}^N\sum_k |\iprod{\gamma_{i,k}^{(l_1)}}{\Phi_{1i}}|^2
\end{split}
\end{equation}
Note that $0\le\sum_k \iprod{\Phi_{2i}}{\gamma_{i,k}^{(l_1)}} \iprod{\gamma_{i,k}^{(l_1)}}{\Phi_{2i}}\le1$.
The upper bound of $n_{1}^{l_1}/d_1$ can be obtained by
\begin{equation}\label{equ:nphiU}
\begin{split}
    &n_{1}^{l_1}/d_1 \le |c_0|^2 n_{2}^{l_1}/d_2 + |c_1|^2 n_{3}^{l_1}/d_3 + |c_2|^2 \cdot N  \\
    &+ 2|c_0 c_1| \sqrt{\frac{n_{2}^{l_1} n_{3}^{l_1}}{d_2 d_3}} + 2|c_0 c_2| \sqrt{\frac{n_{2}^{l_1} N}{d_2}}
    + 2|c_1 c_2| \sqrt{\frac{n_{3}^{l_1} N}{d_3}}
\end{split}
\end{equation}
and we also have the lower bound
\begin{equation}\label{equ:nphiL}
\begin{split}
    &n_{1}^{l_1}/d_1 \ge |c_0|^2 n_{2}^{l_1}/d_2 + |c_1|^2 n_{3}^{l_1}/d_3    \\
    &- \left[ 2|c_0 c_1| \sqrt{\frac{n_{2}^{l_1} n_{3}^{l_1}}{d_2 d_3}} + 2|c_0 c_2| \sqrt{\frac{n_{2}^{l_1} N}{d_2}}
    + 2|c_1 c_2| \sqrt{\frac{n_{3}^{l_1} N}{d_3}} \right]
\end{split}
\end{equation}
Then we define the yield of state $\ket\phi$ for outcome $l_1$ by
\begin{equation}\label{equ:Sphi}
    S_\phi^{l_1} = \frac{n_{1}^{l_1}}{d_1 N}
\end{equation}
where
$d_1 N$ is the total number of time windows that use source 1, which emits  the real-photon state $|\phi\rangle$. And similarly
\begin{equation}\label{equ:Sphi0}
    S_{\phi_0}^{l_1} = \frac{n_{2}^{l_1}}{d_2 N},\ S_{\phi_1}^{l_1} = \frac{n_{3}^{l_1}}{d_3 N}
\end{equation}

Therefore, Eqs. (\ref{equ:nphiU})(\ref{equ:nphiL}) can be written in the form of yield:
\begin{equation}\label{equ:SphiU}
\begin{split}
    S_{\phi}^{l_1} \le& |c_0|^2 S_{\phi_0}^{l_1} + |c_1|^2 S_{\phi_1}^{l_1} + |c_2|^2  \\
    &+ 2|c_0 c_1| \sqrt{S_{\phi_0}^{l_1} S_{\phi_1}^{l_1}} + 2|c_0 c_2| \sqrt{S_{\phi_0}^{l_1}}
    + 2|c_1 c_2| \sqrt{S_{\phi_1}^{l_1}}
\end{split}
\end{equation}
and
\begin{equation}\label{equ:SphiL}
\begin{split}
    S_{\phi}^{l_1} \ge& |c_0|^2 S_{\phi_0}^{l_1} + |c_1|^2 S_{\phi_1}^{l_1}    \\
    &- \left[ 2|c_0 c_1| \sqrt{S_{\phi_0}^{l_1} S_{\phi_1}^{l_1}} + 2|c_0 c_2| \sqrt{S_{\phi_0}^{l_1}}
    + 2|c_1 c_2| \sqrt{S_{\phi_1}^{l_1}} \right]
\end{split}
\end{equation}
With Eqs. (\ref{equ:SphiU})(\ref{equ:SphiL}), {\em they} can estimate the bounds of yield of state $\ket{\phi}$ for $l_1$-event even  using the observed data of other states.
\subsubsection{Bound-value estimation in the protocol}
In the protocol, we use real state $|0,0\rangle$ in a time window $Z_{\mathcal O}$ and real state $|\alpha_A,\alpha_B\rangle$ in a time window $Z_{\mathcal B}$.
The yield for $L$-event or $R$-event of this two kinds of time windows can be observed directly from the time windows of random subset of $v$.
We need to use these observed data to calculate bound values for fraction of time windows heralded by the outcome $l_1$  among all $X_+$-windows, say $S_{X_+}^{l_1}$. In particular the lower bound for $L$-event $\underline{S}_{X_+}^L$  and
the upper bound for $R$-event $\bar{S}_{X_+}^R$.
We can directly apply Eqs. (\ref{equ:SphiU})(\ref{equ:SphiL}), with the real-photon states $|\phi\rangle$, $|\phi_0\rangle$, and $|\phi_1\rangle$
 being replaced by $|\chi^+\rangle$, $|0,0\rangle$, and $|\alpha_A,\alpha_B\rangle$, respectively.
We can easily to obtain:
\begin{equation}\label{equ:chi+}
    \ket{\chi^+} = \frac{e^{-\mu}\ket{0,0} + \ket{\alpha_A,\alpha_B} - (1-e^{-\mu})\ket{\tilde\alpha_A,\tilde\alpha_B}}{e^{-\mu/2}\sqrt{2(1+e^{-\mu})}}
\end{equation}
remind that $\sqrt{1-e^{-\mu}}|\tilde \alpha_x\rangle = |\alpha_x\rangle - e^{-\mu/2}|0\rangle,x=A,B$. Therefore,
\begin{equation}
\begin{split}
    c_0 &= \frac{e^{-\mu/2}}{\sqrt{2(1+e^{-\mu})}} \\
    c_1 &= \frac{1}{e^{-\mu/2}\sqrt{2(1+e^{-\mu})}} \\
    c_2 &= \frac{1-e^{-\mu}}{e^{-\mu/2}\sqrt{2(1+e^{-\mu})}} \\
\end{split}
\end{equation}
 We use notation $S_y^{l_1}$ for the fraction of effective windows heralded by outcome ${l_1}$ among all $y$-windows of the test set $v$ and ${l_1}=L,R$,  $y={X_+},Z_{\mathcal B},Z_{\mathcal O}$.
 Say, if there are $n^{l_1}_{y}$ effective windows
 \begin{equation}\label{equ:SX+U}
\begin{split}
    &S_{X_+}^{l_1} \le \overline{S}_{X_+}^{l_1} = \frac{1}{2(1+e^{-\mu})} \{ e^{-\mu} S_{\mathcal O}^{l_1} + e^{\mu} S_{\mathcal B}^{l_1} + \frac{(1-e^{-\mu})^2}{e^{-\mu}} \\
    &+  2\sqrt{S_{\mathcal O}^{l_1} S_{\mathcal B}^{l_1}} + 2(1-e^{-\mu}) \sqrt{S_{\mathcal O}^{l_1}} + \frac{2(1-e^{-\mu})}{e^{-\mu}} \sqrt{S_{\mathcal B}^{l_1}} \}
\end{split}
\end{equation}
\begin{equation}\label{equ:nX+L}
\begin{split}
    &S_{X_+}^{l_1} \ge \underline{S}_{X_+}^{l_1} = \frac{1}{2(1+e^{-\mu})} \{ e^{-\mu} S_{\mathcal O}^{l_1} + e^{\mu} S_{\mathcal B}^{l_1} \\
    &- 2[\sqrt{S_{\mathcal O}^{l_1} S_{\mathcal B}^{l_1}} + (1-e^{-\mu}) \sqrt{S_{\mathcal O}^{l_1}}
    + \frac{1-e^{-\mu}}{e^{-\mu}} (\sqrt{S_{\mathcal B}^{l_1}})] \}
\end{split}
\end{equation}
 Replacing $l_1$ in Eq.(\ref{equ:SX+U})  above by $R$ and $l_1$ in Eq.(\ref{equ:nX+L})  by $L$, we obtain  $\bar S_{X_+}^R$ and $\underline{S}_{X_+}^L$.
\subsubsection{Estimate the bound of the phase-flip error rate.}
Suppose set $u$ consists of $M$ $X$-windows, among which there are  $(1+e^{-\mu})M/2$ $X_+$-windows and  $(1-e^{-\mu})M/2$ $X_-$-windows.
The phase-flip error rate is defined as
\begin{equation}\label{equ:eph}
    e^{ph} = \frac{n_{X_+}^{R} + n_{X_-}^{L}}{n_{X_+}^{L} + n_{X_+}^{R} + n_{X_-}^{L} + n_{X_-}^{R}}
\end{equation}
where $n_a^d$ represent the number of effective $a$-windows heralded by detector $d$ clicking. $d=L,R$ and $a={X_+},{X_-}$.
Straightly, we have
\begin{equation}\label{equ:c2}
\begin{split}
    e^{ph} &= \frac{n_{X_+}^{R}-n_{X_+}^{L}+n_X^{L}}{n_X} \\
    &= \frac{(1+e^{-\mu}) \left[S_{X_+}^{R}-S_{X_+}^{L}\right] +2S_{X}^{L}} {2S_X}
\end{split}
\end{equation}
where $n_X$ and $S_X$ are the number of effective events happened in $X$-windows and the fraction of effective windows among all $X$-windows, respectively.
We have the following formula for the upper bound of phase-flip error rate
\begin{equation}\label{equ:c3}
    e^{ph}\le \overline{e}^{ph} = \frac{(1+e^{-\mu}) \left[\overline{S}_{X_+}^{R}-\underline{S}_{X_+}^{L}\right] +2S_{\tilde Z}^{L}} {2S_{\tilde Z}}
\end{equation}
Here,  $S_{\tilde Z}$ is the fraction of effective windows among those $\tilde Z$-windows in set $u$, which plays the role of fake $X$-windows in the real protocol.
$S_{\tilde Z}^{L}$ is the fraction of effective windows heralded by the left detector among all $\tilde Z$-windows in set $u$. Since a $\tilde Z$-window in set
$u$ is identical to a $\tilde Z$-window in set $v$, {\em they} can obtain the values of $S_{\tilde Z},S_{\tilde Z}^{L}$ by directly observing
set $v$.   The quantities of $\overline{S}_{X_+}^{R}$ and $\underline{S}_{X_+}^{L}$ can be calculated by Eqs.(\ref{equ:SX+U},\ref{equ:nX+L}).

\end{document}